\documentclass[preprint]{aastex6}
\usepackage{graphicx}
\usepackage{epsfig}
\usepackage{multirow}
\usepackage{footnote}

\shorttitle{Abundances in the Polluted White Dwarf SDSS\,J1043+0855}
\shortauthors{C. Melis \& P. Dufour}

\begin{document}


\title{\large \bf Does a Differentiated, Carbonate-Rich, Rocky Object Pollute the White Dwarf SDSS\,J104341.53+085558.2?}


\author{\large Carl Melis\altaffilmark{1} and P. Dufour\altaffilmark{2}}
\affil{email: cmelis@ucsd.edu \\
$^1$Center for Astrophysics and Space Sciences, University of California, San Diego, CA 92093-0424, USA \\
$^2$Institut de Recherche sur les Exoplan\`{e}tes (iREx), Universit\'{e} de Montr\'{e}al,
Montr\'{e}al, QC H3C 3J7, Canada \\
}










\begin{abstract}

\large{
We present spectroscopic observations of the dust- and gas-enshrouded, polluted,
single white dwarf star SDSS\,J104341.53+085558.2 (hereafter SDSS\,J1043+0855). 
$Hubble$ $Space$ $Telescope$ Cosmic Origins Spectrograph far-ultraviolet spectra 
combined with deep Keck HIRES optical spectroscopy reveal the elements 
C, O, Mg, Al, Si, P, S, Ca, Fe, and Ni and enable useful limits for Sc, Ti, V, Cr, 
and Mn in the photosphere of SDSS\,J1043+0855. From this suite of elements we 
determine that the parent body being accreted by SDSS\,J1043+0855 is
similar to the silicate Moon or the outer layers of Earth in that it is rocky and iron-poor. 
Combining this with comparison to other heavily polluted white dwarf stars, we are able to
identify the material being accreted by SDSS\,J1043+0855 as likely to have come from 
the outermost layers of a differentiated object.
Furthermore, we present evidence that some polluted white dwarfs (including SDSS\,J1043+0855)
allow us to examine the structure of differentiated extrasolar rocky bodies.
Enhanced levels of carbon in the body polluting SDSS\,J1043+0855
relative to the Earth-Moon system can be explained with a model where a significant amount 
of the accreted rocky minerals took the form of carbonates; specifically, through this model
the accreted material could be up to 9\% calcium-carbonate by mass.
}

\end{abstract}




\large{

\section{\large \bf Introduction}


Exoplanet surveys have begun to identify and characterize rocky
planets with increasing frequency.
As these attempts push closer to routine detection of Earth-like planets
in terms of mass, radius, and orbital characteristics \citep[e.g.,][]{burke15}, it is natural to wonder
just how Earth-like these other worlds are in terms of composition.
With only masses and radii available for them, most exoplanets cannot
have well-constrained compositions inferred 
and setting limits on their internal structure is challenging at best
\citep[e.g.,][and references therein]{dorn15,rogers15,vanlieshout16}.
Theoretical efforts provide some insight into how different chemical
compositions in a protoplanetary disk can give rise to a range of rocky
planet bulk compositions and internal structures
and in some cases provide interesting comparisons to observations
\citep[e.g.,][and references therein]{bond10,carterbond12a,carterbond12b,elser12,moriarty14,thiabaud14,thiabaud15,pignatale16,unterborn16b,alessi16}

While conventional exoplanet search methods are not yet able to arrive at definitive conclusions 
about the interior structure and bulk composition of exoplanets, studies of white dwarf stars
have been providing the bulk chemical composition 
for rocky bodies outside of the Solar system in unprecedented detail 
(e.g., rare elements such as strontium and scandium $-$ both down in
abundance from hydrogen in the Sun by a factor of a billion $-$ can be detected
in the atmosphere of the most heavily polluted white dwarfs) 
since the seminal publication on the topic almost a decade ago \citep{zuckerman07}.
Since that time, new discoveries and capabilities have cemented the connection
between disrupted rocky bodies and the pollution in the atmosphere of
white dwarf stars
\citep[e.g.,][and references therein]{vanderburg15,xu16,gaensicke16,alonso16,zhou16,rappaport16}.
These works $-$ combined with the library of previous research in the field of
dusty, polluted, single white dwarf stars
\citep[e.g.,][and references therein]{debes02,jura03b,jura08,farihi09,veras16} $-$
show that white dwarf stars regularly accrete material 
from their extant planetary systems and that we can use them to
characterize the composition of the accreted rocky bodies.

Enough polluted white dwarf stars are now known and sufficiently well-characterized
that synthesis of these objects as a population, and the resulting insight into
planets or planetesimals in the local Galaxy, can be made. \citet{jura06wd} 
and \citet{wilson16} show that extrasolar
planetesimals are carbon-deficient, much like the inner planetary system of our
own Solar system or similar to primitive chondritic material. 
\citet{juraxu12} provide insight into the prevalence of water-rich
rocky bodies, showing that as a population these objects are dry, $<$1\% water
by mass \citep[but see][]{farihi13,raddi15,farihi16} while \citet{jura13a}  
similarly show that refractory-rich planetesimals are unlikely to be accreted by
white dwarf stars. \citet{jura13b} 
suggest that radiogenic heating of rocky bodies through $^{26}$Al may be common
and thus that our Solar system is not unique in having significant quantities of this
nuclide. \citet{jura14} 
present a search methodology that can place interesting limits on evidence for plate
tectonics in differentiated extrasolar rocky objects.

In this paper we present a thorough ultraviolet and optical spectroscopic inventory
of the material polluting the atmosphere of the dust- and gas-disk hosting single
white dwarf star SDSS J104341.53+085558.2 (hereafter SDSS\,J1043+0855).
We find this material to be rocky in general, extremely depleted in iron, and
enhanced in carbon relative to rocky material in the inner parts of our Solar system. 
A synthesis of all heavily polluted white dwarf star measurements is made, resulting in
a strong suggestion that some white dwarfs are experiencing pollution from specific
regions of differentiated rocky bodies, thus opening a window into extrasolar rocky
body structure.
Carbonates are proposed as a possible carrier of the enhanced carbon signature
for the parent body polluting SDSS\,J1043+0855 and the implications of such a
scenario are discussed.

\section{\large \bf Observations}
\label{secobs}




\subsection{\large Ultraviolet Spectroscopy}
\label{secuvspec}

$Hubble$ $Space$ $Telescope$ Cosmic Origins Spectrograph (hereafter COS; 
\citealt{green12}) observations were obtained for SDSS\,J1043+0855 on 
2015 April 27. Two spacecraft orbits were awarded to this target as part of GO program 13700 
and we obtained a total exposure time of 5105~seconds with the G130M 
grating centered at a wavelength of 1291\,\AA . Observations were obtained in the {\sf TIME-TAG} 
mode using the 2.5$''$ diameter primary science aperture. Four different {\sf FP-POS} values
were utilized during the observations (about a quarter of the total time was spent in each)
to minimize fixed pattern noise in the detector.
This setup yields spectral resolving power of $\approx$20,000
and final spectral coverage from 1133 to 1433\,\AA\
except for a gap from 1279 to 1288\,\AA .

The raw COS data were processed with the {\sf \verb+CALCOS+} pipeline 3.0,
coadded with the use of the IDL script {\sf \verb+COADD_X1D+}
\citep{danforth10,keeney12}, and 
then smoothed with a boxcar of six pixels. The resulting signal-to-noise-ratio (S/N) per pixel
in the smoothed data is $\approx$14 at a wavelength of 1320\,\AA . 
The final spectrum presented here is flux calibrated and in vacuum wavelengths corrected to the
heliocentric reference frame. 
Following \citet{jura12}, we use the {\sf timefilter} module to extract the 
night-time portion of the data around the O~I lines between 1300 and 1308\,\AA\
to help mitigate terrestrial day airglow emission. This is mostly done
as a sanity check on the abundance 
for the one uncontaminated O~I line near 1152\,\AA\ which otherwise
would be the only line to inform the oxygen abundance in SDSS\,J1043+0855.
Lower S/N spectra are obtained for the O~I lines between 1300 and 1308\,\AA\
as a result of the filtering and abundances consistent with the 1152\,\AA\ line are obtained.

\subsection{\large Optical Spectroscopy}
\label{sechobs}

\citet{melis10} provide a detailed accounting of
Keck HIRES \citep{vogt94} optical echelle spectra that were obtained for SDSS\,J1043+0855. 
Optical observations obtained by \citet{melis10}
had an aim of exploring the gaseous component of the debris disk
orbiting the white dwarf star. In this work we re-reduce the HIRES data with a focus
on making measurements or obtaining tight limits for metal absorption lines in the spectra.
Data were reduced using the
{\sf MAKEE} software package which outputs heliocentric velocity corrected spectra
shifted to vacuum wavelengths. After reduction and extraction, 
high order polynomials are fit to each order to bring overlapping 
order segments into agreement before combining all orders of every
HIRES exposure to generate a final spectrum for analysis.

\section{\large \bf Results and Modeling}
\label{secspec}

We adopt SDSS\,J1043+0855 stellar parameters from \citet{tremblay11}:
a DA white dwarf with effective temperature of
18,330\,K, log$g$ of 8.05 (cgs), resulting mass of 0.626\,M$_{\odot}$, and
a ``thin envelope'' having mass ratio $q$$=$M$_{\rm env}$/M$_{\rm WD}$$=$10$^{-16.645}$.
We note that these parameters are, to within quoted errors in each source paper,
consistent with those given by \citet{manser16}.
With these parameters we proceed in fitting the metallic absorption lines detected 
in the COS and HIRES spectra and deriving upper limits for elements of interest. 
Fitting proceeds very similar as to the description given in \citet{melis11} and references therein.
We use a local thermodynamic equillibrium (LTE) 
model atmosphere code similar to that described in 
\citet{dufour05,dufour07} and 
absorption line data are taken from the Vienna Atomic Line Database.

We calculate grids of synthetic spectra for each element of interest
covering a range of abundances typically from log[$n$(Z)/$n$(H)]= $-$3.0 to
$-$8.0 in steps of 0.5 dex. We then determine the abundance of each
element by fitting the various observed lines using a similar method
to that described in \citet{dufour05}. Briefly, this is done by
minimizing the value of $\chi$$^2$ taken as the sum over all frequencies
of the difference between the normalized observed and model fluxes
(the synthetic spectra are multiplied by a constant factor to account for the solid angle 
and the slope of the spectra locally are allowed to vary by a first order polynomial
to account for residuals from the normalization procedure), 
all frequency points being given an equal weight. 
Interpolation between grid points allows us to achieve individual line abundances accurate to $<$0.05 dex. This is done individually for each line and the final adopted
abundances (see Table \ref{tab1043}) are taken to be the average of all the
measurements made for a given element. 
Uncertainties were taken to be the dispersion among abundance
values used in the average. If this dispersion was less than 0.20 dex, then
the abundance uncertainty is set to 0.20 dex. For elements where only
one line is available to derive the abundance, the uncertainty is set
to 0.30 dex.
It is noted that changing the effective temperature by $\pm$500\,K 
and/or the surface gravity by $\pm$0.1~dex
(typical of uncertainties for white dwarf stellar atmospheric modeling)
changes derived abundances by less than 0.1~dex, indicating that the relative abundance 
ratios are not sensitive to the exact final stellar parameters adopted.

We find contributions from C, O, Al, Si, P, S, Fe, and Ni in the COS data
and Ca, Mg, and Si in the HIRES data (Figure \ref{figmetals}). Additionally,
we derive from COS data restrictive upper limits for Sc, Ti, V, Cr, and Mn (see Table \ref{tab1043})
and from HIRES data a non-restrictive upper limit for log[n(Na)/n(H)] of $<$10$^{-5}$
(this value is not quoted in the table).
Abundances for all elements identified in the atmosphere
of SDSS\,J1043+0855 are reported in Table \ref{tab1043} as well as upper limits
for other elements of interest (see Section \ref{secdisc}).
S is a tentative detection and the suggested abundance can be taken as a firm upper limit;
higher S/N data can better reveal this element.
We note that, similar to \citet{gaensicke12} and references therein, we find a 
discrepancy between the abundance for Si as derived from optical and ultraviolet
data (see discussion in \citealt{gaensicke12} for reasons why these values may
not agree). We adopt here the average of the optical and ultraviolet Si abundances
and set the error to cover the difference between them (Table \ref{tab1043}).

Interstellar lines are present in the optical and ultraviolet spectra, particularly near
the ultraviolet C lines. For C, these lines are readily identifiable by their reversed line strength ratio
due to the material being in a dramatically different physical environment than the
atmosphere of the white dwarf star. All interstellar lines are also 
offset in radial velocity relative to the
stellar photospheric velocity. \citet{melis10} measure a photospheric velocity
(which includes gravitational redshift) of +39$\pm$4\,km\,s$^{-1}$, a value
reproduced within the uncertainties by all new photospheric elements presented
herein. Interstellar lines, specifically C and Ca, have velocity near 0\,km\,s$^{-1}$
(see also \citealt{melis10}). We note the excellent agreement of optical abundance
values for Ca, Mg, and Si with those reported in \citet{manser16}.

\section{\large \bf Discussion}
\label{secdisc}

Previous work has shown that SDSS\,J1043+0855 is host to both
gaseous and dusty debris within its tidal radius
\citep{gaensicke07,melis10,brinkworth12,manser16}, suggesting that any
metal pollution seen in its atmosphere originated in the tidal
disruption and subsequent accretion of a rocky object from the
white dwarf's extant planetary system (e.g., \citealt{jura03b}).
Following this line of logic, we can estimate the elemental composition
of the material currently being accreted by examining the
metal abundances reported in Table \ref{tab1043}.
Similar to the case of the hot white dwarf GALEX\,J193156.8+011745
as described in \citet{melis11}, we expect that SDSS\,J1043+0855
is in the steady-state accretion phase and that radiative levitation
is negligible (see also \citealt{xu13,xu14}). We then
use Equation 16 from \citet{jura09b} and diffusion constants
calculated by G.\ Fontaine\footnote{http://dev.montrealwhitedwarfdatabase.org/evolution.html}
(private comm., 2016; see Table \ref{tab1043} and \citealt{dufour16})
with the parameters for SDSS\,J1043+0855
to estimate the accretion rate and abundance ratio relative to Mg
of the polluting material for each element (Si is not used for this comparison because
of the high adopted uncertainty for its abundance $-$ see Section \ref{secspec}).

Accreted abundances by number 
relative to Mg and mass accretion rates are reported in Table \ref{tab1043}.
Distinct amongst these are high levels of carbon ($\approx$1\% by mass) and calcium
($\approx$7\% by mass) and deficient iron ($\approx$10\% by mass). The
significant deficit of iron is reminiscent of the abundances for the surface of the Moon
(hereafter the silicate Moon) or for the crust and mantle of the Earth 
(compare with Figure 7 of \citealt{jurayoung14}). To help illustrate
this, Table \ref{tab1043} also reports metal abundance ratios relative to magnesium for
the silicate Moon as taken from Table 4.7 of \citet{lodders11} with the exception of
carbon which comes from \citet{wetzel15}. Comparison between the silicate Moon and
measurements for SDSS\,J1043+0855 are also plotted in Figure \ref{fig1043z}. Also
displayed in Figure \ref{fig1043z} is a ``Hollow Earth'' model. This model is constructed
in a similar fashion as the ``wind-stripped Earth'' model presented in Figure 6 of
\citet{melis11}, except it starts removing material from the core and works its way out toward
the crust. We remove varying fractions of the core, then lower mantle, then upper mantle
and compute the resulting bulk abundances of the remaining material relative to magnesium.
The best match with the data for SDSS\,J1043+0855 is shown in Figure \ref{fig1043z}
and corresponds to removing the entire core and lower mantle of the Earth
(this is similar to the case of NLTT\,43806; see \citealt{zuckerman11} and discussion below).

Based on the comparison shown in Figure \ref{fig1043z} it would seem plausible
that SDSS\,J1043+0855 is accreting the surface layers of either an exomoon
(e.g., \citealt{payne16}) or
fully-differentiated massive rocky body. The agreement between enhancements in nickel
and calcium relative to the silicate Moon for both the Hollow Earth model and SDSS\,J1043+0855
possibly suggest better agreement with the latter scenario, although this relies on
strong assumptions about commonality between the parent body feeding SDSS\,J1043+0855
and the Earth (an assumption clearly not borne out by elements like aluminum and carbon);
a way to further test this idea would be through tighter limits on manganese, chromium, and
especially titanium (which is strongly depleted for the surface layers of the Earth).

Corroboration of a massive differentiated body as an appropriate interpretation
could come from limits on the total mass of the accreted body. However, from the
total mass accretion rate of $\approx$1.3$\times$10$^{8}$\,g\,s$^{-1}$ and under
the assumption of this being constant over a disk lifetime of 
$\sim$1\,Myr \citep{melis11,girven12}, we arrive
at an estimate for the total initial mass of the rocky material of $\sim$4$\times$10$^{21}$\,g. 
This is well below the estimated total mass of any planet or major moon in the Solar system.
But, as discussed earlier in this section and as outlined in detail below,
the body being accreted by SDSS\,J1043+0855 very likely was differentiated
and as such should have been at least as massive as objects like Vesta
or Ceres in our Solar system (10$^{23}$-10$^{24}$\,g;
as far as we know, these are the least massive differentiated bodies in the Solar system).
As such, we should view the total mass given above as a lower limit; indeed, the assumption
that the currently observed accretion rate would be constant over the entire disk lifetime
is suspect as several studies find instead evidence that greatly enhanced accretion
rates can be realized during the evolution of a white dwarf debris disk
\citep[e.g.,][]{rafikov11a,rafikov11b,girven12,farihi12,wyatt14}. Alternatively, and as
described in more detail below, it could be the case that only a small fraction of rocky
material was liberated from a differentiated parent body and subsequently delivered
to SDSS\,J1043+0855 (see also \citealt{zuckerman11}).

Simulation work suggests that it is unlikely that a full-fledged planet 
(Mars-like or more massive) would be delivered
to the tidal destruction radius of a white dwarf star
(e.g., \citealt{debes02}; \citealt{jura08}; \citealt{mustill14}; \citealt{veras15}; \citealt{veras16}).
As discussed by \citet{zuckerman11}, it is not necessary to deliver an entire planet-sized
object into the tidal destruction radius of a polluted white dwarf star, 
even when there is evidence for the material to have come from particular
layers of a differentiated rocky body.
One need only to liberate some amount of material from the relevant 
region that is believed to now be polluting the white dwarf star under study. Having just 
some pieces removed from a differentiated rocky body $-$ especially surface material $-$ is 
rather plausible as evidenced by simulations of giant impact type collisions during rocky 
planet formation (e.g., \citealt{asphaug14} and references therein; \citealt{melis11} describes
how deeper layers may become more readily exposed before impact liberation occurs).
Indeed, such collisions could very well generate material with an orbit that 
would take it to where it would be gravitationally captured and eventually accreted by the 
host white dwarf star. The situation with moons that are much smaller than 
Mars could be somewhat different as it is probably possible to get such objects in whole into 
their host white dwarf's tidal destruction radius (e.g., \citealt{payne16}).

Perhaps most striking in Figure \ref{fig1043z} is the dramatic enhancement of carbon
in the parent body being accreted by SDSS\,J1043+0855 relative to the silicate Moon
and the ``Hollow Earth'' model. It is worth noting that measurements of carbon in the
Moon and for the surface layers of the Earth have produced a wide range of results
in the past with values as much as two orders of magnitude larger than that reported
in Table \ref{tab1043}. We use the value from \citet{wetzel15} who have made 
significant improvements over previous attempts.
As such, while their measurement is the state-of-the-art,
it bears mentioning that measurements of carbon for
the Moon and surface layers of Earth have a shaky past.
Regardless, the enhancement of carbon in the body polluting SDSS\,J1043+0855
relative to the surface of the Earth and the Moon will likely remain (even if at some lower level).

White dwarfs that are accreting rocky material $\gtrsim$1\% carbon by mass have been
previously reported \citep[e.g.,][]{xu13,jura15,wilson15,wilson16}, although each of these are
helium-dominated atmosphere white dwarfs while SDSS\,J1043+0855 is
hydrogen-dominated. The importance of this distinction
is discussed in \citet{wilson15} and summarized here:
helium-atmosphere white dwarfs are susceptible to dredge-up from the interior of 
the white dwarf star which can implant sufficient carbon to explain observations to date.
A hydrogen-atmosphere white dwarf, especially at the temperature of SDSS\,J1043+0855,
would not have atmospheric carbon pollution as a result of such a mechanism.
SDSS\,J1043+0855 could thus be considered the first high-carbon content
polluted white dwarf where the carbon is unambiguously
delivered by outside sources. However, with this designation comes 
the inevitable conclusion that other such white dwarfs are probably similar.

Something that is thus far unique to the case of SDSS\,J1043+0855 is the C/O ratio realized for
its polluting material. While carbon does not dominate over oxygen, the C/O ratio does
lie squarely in between the ``bi-modal'' distribution suggested for bodies polluting white
dwarf stars in \citet{wilson16}. This suggests that there is indeed more diversity within
exoplanetary rocky bodies yet to be discovered and that distinctly non-Solar system-like
material does seem to exist in the local Galaxy, although it does appear to be rare.
It is desirable to know how the carbon was originally delivered to the rocky body 
that is now being accreted by SDSS\,J1043+0855 and how it populated the rocky body 
such that it remained entrained through to the white dwarf
phase of stellar evolution. Below we explore a plausible idea for the retainment of the carbon
that is thus far consistent with the data available.

Following \citet{klein10,klein11} and \citet{farihi13}, we attempt to balance the accreted 
oxygen abundance under the expectation that all oxygen was contained
in rocky oxygenated minerals (e.g., MgO, Al$_2$O$_3$, SiO$_2$, CaO, 
FeO, and NiO). In this way, we find that $-$ within the errors $-$ oxygen is 
roughly critically distributed into rocky minerals (a small oxygen excess results from
inclusion of just the above mentioned metal oxides). 
Some forms of carbon (e.g., CO$_2$) might be lost in a manner similar to water 
during the host star's asymptotic giant branch phase of evolution \citep[see][]{jura10}.
Suppose that the carbon in the parent body now being accreted by SDSS\,J1043+0855
was instead locked up in carbonate minerals and as such would be less 
susceptible to loss mechanisms.
Of the elements seen thus far in spectroscopic observations of SDSS\,J1043+0855,
we could construct the following naturally occurring carbonates (on Earth): 
CaCO$_3$, MgCO$_3$, and FeCO$_3$.
CaCO$_3$ in particular has already been seen in various space environments
(e.g., \citealt{kemper02a}; \citealt{kemper02b}; \citealt{chiavassa05}; \citealt{boynton09}; \citealt{desanctis16}), 
and thus is a reasonable candidate for also being present in the material
being accreted by SDSS\,J1043+0855.

We examine two possibilities: (i) a generic carbonate scenario where we just account for
how much CO$_2$ would be necessary and what the impact would be on the oxygen budget, and
(ii) a scenario where all carbon is locked up in CaCO$_3$. Both scenarios will have a similar
impact on the oxygen budget, but in the second scenario we can make an estimate for what
the maximum fraction by mass the parent body could have been in calcium-carbonate.
We find in general that allowing carbonates in the oxygen budget calculation 
results in a $\approx$3\% overtaxing of the available oxygen (note that 
it is assumed that all Fe is in the form of FeO which maximizes its contribution to the oxygen budget). 
If no carbonates are allowed in the oxygen
budget calculation, then an oxygen excess of $\approx$5\% is instead obtained.
The maximum mass fraction of CO$_2$ that
can be realized in scenario (i) is $\approx$4\%. In scenario (ii), the maximum mass fraction
of CaCO$_3$ that can be realized is $\approx$9\%.
The result of the oxygen balance analysis is that it appears as though the material being accreted by
SDSS\,J1043+0855 very likely originated in rocky minerals, is only
a small percentage (if any) of water by mass (see e.g., \citealt{farihi13}), 
and could host a significant quantity of
carbonates $-$ especially calcium-carbonate $-$ by mass. 

It is noted that cometary bodies in the Solar system are known to contain 
$\approx$5\% CO$_2$ by mass (e.g, \citealt{jessberger89}),
very similar to the scenario outlined above for the carbon and oxygen excess in the
accreted material polluting SDSS\,J1043+0855. However, a cometary origin for the
material polluting this white dwarf star is not very likely for at least the following reasons:
(i) comets can be up to 80\% water by mass, whereas the object being accreted by 
SDSS\,J1043+0855 is at best equal parts water and CO$_2$ by mass. 
Having a comet that somehow loses all its water but none of its CO$_2$ is improbable,
especially if one considers the loss mechanisms outlined in \citet{jura10}. 
(ii) The otherwise good agreement of the abundances measured in the atmosphere of
SDSS\,J1043+0855 with Earth's crust would be hard to reconcile given that comets
appear rather primitive in their rocky elements (e.g., \citealt{jessberger89}).
Thus, we reject extrasolar comet-like bodies as a viable candidate for explaining the
material polluting SDSS\,J1043+0855.

White dwarfs that are iron-poor or that otherwise have Earth surface-layer-like compositions
have been previously reported \citep[e.g.,][]{zuckerman11,farihi13,raddi15}.
Having securely identified four examples of such systems now with the characterization
of SDSS\,J1043+0855, it seems prudent to evaluate how such systems compare
to other well-characterized polluted white dwarfs. 
We do this by collecting relative abundances and mass fractions of accreted elements
for all available white dwarf stars. We only consider those objects where each of
iron, magnesium, silicon, and oxygen have been measured. There are a few exceptions:
GD\,362 where limits on oxygen are sufficiently deep to robustly estimate its contribution
to the total mass of the accreted body (e.g., \citealt{xu13}), and PG\,1225$-$079 and
NLTT\,43806 where estimates of the oxygen mass fraction are made assuming all metals
$-$ excluding and including iron, respectively $-$ originate in metal oxides (see below).

Examining this collection reveals
a trend in accreted abundance ratios for Si/Fe vs Mg/Fe that is
suggestive of whether accreted material originates from the inner or 
outer regions of a rocky body (Figure \ref{figtrend}). 
Fe-poor objects in the ``Crust-like'' area of Figure \ref{figtrend}
are strongly suggestive of having an origin in the outer-most layers
of a differentiated rocky body. The majority of the Fe in these objects got locked up deep in their
interiors as a result of the differentiation process and only the surface crust and mantle
material was delivered to the host white dwarf. 
Fe-rich objects in the ``Core-like'' area of the figure instead resemble objects whose
outer-most layers were removed and only the inner parts were delivered to the white
dwarf (e.g., \citealt{melis11}). Verification that these bodies truly are iron-poor or iron-rich
as suggested by their placement in the ``Crust-like'' or ``Core-like'' regions of Figure \ref{figtrend}
comes from calculating the mass fraction of iron in the parent body material being accreted
in each white dwarf system (bottom panel of Figure \ref{figtrend}). 

This overall trend in heavily polluted white dwarf stars provides not only clues as to what the
evolutionary history of accreted material is, but also strongly suggests that white dwarf stars
can be used to examine the chemical composition of {\it specific regions} of an extrasolar
rocky body in specialized cases. An exciting prospect of
this interpretation is its potential to inform us on the diversity of the crust-mantle regions 
(i.e., surfaces) of
alien worlds and its implications for the prevalence of water-rich $-$ vastly more so than
the Earth $-$ differentiated rocky objects. 
NLTT\,43806, one of the four systems found in the ``Crust-like'' area of Figure \ref{figtrend},
is distinctly Earth-like in its high mass fraction of aluminum \citep{zuckerman11}. It is 
worth noting that all other ``Crust-like'' white dwarfs fall below the Al/Mg ratio of the 
silicate Moon similar to SDSS\,J1043+0855 in Figure \ref{fig1043z}; whether this apparent
bi-modality in Al-content of ``Crust-like'' white dwarfs continues to manifest as more objects
are discovered and characterized will be interesting as Al is thought to be tied to melt 
processes in the surface of the Earth (J.\ Day, private comm. 2016).
GD\,61 and SDSS\,J1242+5226, the two remaining ``Crust-like'' objects, accrete rocky bodies
which both show evidence for significant water pollution.

Recognizing that the outer layers of the
Earth amount to only about 10\% of the total mass of the planet and given the total mass estimates
for the parent bodies polluting ``Crust-like'' white dwarfs, 
it is plausible to interpret material polluting these white dwarf stars as having its
origin in the surface layers of a full-fledged planet (see also discussion in \citealt{zuckerman11}).
If correct, this interpretation would suggest that out of four examples of
``Crust-like'' rocky material sampled to date, one is distinctly Earth-like, two are extremely
water-rich, and one shows enhanced carbon that could plausibly have
had its origin in carbonates. While clearly within the
realm of statistics of small numbers, these possibilities surely warrant deeper
investigation and a broader search for similarly iron-poor polluted white dwarf stars.



It is of some value to
consider the implications of SDSS\,J1043+0855 as being polluted by the outer
layers of a massive, differentiated rocky body rich in carbonates, specifically calcium-carbonate. 
In general, carbonates on Earth and other space environments
(including, but not limited to: Ceres, Mars, protoplanetary disks, evolved stars, and planetary nebulae;
see \citealt{desanctis16}, \citealt{boynton09}, \citealt{chiavassa05}, \citealt{kemper02b},
\citealt{kemper02a}, respectively)
typically require some amount of water to form (e.g., \citealt{toppani05}). To 
produce on the order of several percent of the mass of
an object's surface in carbonates would thus suggest a reasonable liquid water
reservoir coupled with some effective CO$_2$-delivery mechanism
(whether that would be in the form of a dense atmosphere, intense outgassing,
or some other mechanism), similar to the conditions
that could have led to the generation of calcium-carbonate in other space environments
\citep{toppani05}.
This argues for the possibility of a liquid water ocean in the outer layers of the parent
body now being accreted by SDSS\,J1043+0855. If true, either this water was removed during
the star's post-main sequence evolution \citep[e.g.,][]{jura10} or is sufficiently small
in mass fraction to be undetectable within our current measurement capabilities
(note that our ocean is $\approx$0.02\% of Earth's mass).

Calcium-carbonate on the Earth is typically associated with marine organisms which 
utilize it as a part of their exoskeletal structures. As mentioned above, calcium-carbonate 
can be seen in a variety of space environments and so it is of course possible to generate
the mineral through non-biological pathways
(but see also the discussion regarding entraining CO$_2$ through the aid of life
by \citealt{haqqmisra16}). Perhaps more vexing for having this mineral originate
through life processes is the extremely short main sequence lifetime of the progenitor
star of SDSS\,J1043+0855 (with a mass of 2.76$\pm$0.04\,M$_{\odot}$ as given
by \citealt{manser16}, this star would have evolved off of the main sequence
in $\lesssim$1\,Gyr). 

Calcium-carbonate in the surface of the world now being accreted by SDSS\,J1043+0855 
could have been synthesized rapidly as the mineral travertine through plate tectonic 
processes that release CO$_2$ and geothermal energy in an aqueous environment. 
Some evidence in favor of such an interpretation already exists.
Recall that the pollution for all four of the ``Crust-like'' 
white dwarfs has distinctly oceanic crust-like signatures of Mg/Fe (Figure \ref{figtrend}). 
The oceanic crust in Earth is regularly polluted by upwellings of magnesium-rich material 
from the mantle through subduction and recycling (plate tectonics processes).
This results in Fe being depleted relative 
to Mg in addition to it being depleted relative to Si as one would expect in the Si-dominated 
continental crust of Earth. Mg and Si in the oceanic crust are comparable and substantially 
more abundant than Fe, while in the continental crust Fe and Mg have comparable 
abundance and are both strongly depleted relative to Si (see Figure \ref{figtrend}).
Additionally, the strong evidence for water in at least two of the ``Crust-like'' 
polluted white dwarfs (and circumstantial evidence for 
water in all of them) suggests atmospheric conditions that allow water to persist through the 
object's life, a characteristic considered prerequisite for tectonic activity 
(consider the comparison between dry Venus which has no plate tectonics 
and the Earth; e.g., \citealt{kaula81}; \citealt{barnes09}).
If this hypothesis pans out, then the results herein
would represent the first evidence for plate tectonics outside
of the Solar system (e.g., see \citealt{jura14}).

While it is fun to speculate about the origin of any carbonates in this object, 
what can be done to confirm that such minerals are actually present? 
The launch of the $James$ $Webb$ $Space$ $Telescope$ is imminent,
and with it will return a capability to perform sensitive mid-infrared spectroscopic
observations \citep{rieke15,kendrew15}. Mid-infrared spectroscopy of SDSS\,J1043+0855 could
detect a characteristic signature of CaCO$_3$ in the disk that orbits it
(e.g., \citealt{kemper02b}), 
thus validating some of the speculative discussion given here. In the meantime,
coupled dynamical and chemical evolution-tracking simulations (such as those in \citealt{bond10})
can explore how one might obtain a rocky body with the chemical abundances exhibited
by that being accreted by SDSS\,J1043+0855.

\section{\large \bf Conclusions}

Spectroscopic observations of SDSS\,J1043+0855 in the ultraviolet
and optical reveal a suite of elements that suggest a parent body
that is rocky and iron-poor. 
Synthesizing all available heavily-polluted white dwarf measurements, we find a 
trend in the accreted abundance ratios for Si/Fe vs Mg/Fe suggestive of whether accreted 
material could have originated from the inner or outer regions of a 
differentiated object; we use this trend to identify the material
polluting SDSS\,J1043+0855 as having its origin in the
outer layers of a differentiated, massive, rocky body.
Enhanced levels of carbon can be explained by a model
where the rocky body polluting SDSS\,J1043+0855 hosted
significant quantities of carbonate minerals $-$ this combined
with a mild enhancement of calcium suggests the possibility
of calcium-carbonate. Future observations with $JWST$ can explore
this interpretation while modeling can provide insight into the formation mechanisms and
evolutionary paths such material might have experienced leading up to its
subsequent tidal destruction and accretion by SDSS\,J1043+0855.


\acknowledgments

C.M.\ would like to acknowledge the profound impact the late Prof.\ Michael Jura 
had on his development
into a scientist; he will be sorely missed, especially within the community who study
white dwarf planetary systems. Indeed,
the present paper benefited significantly from discussions with Prof.\ Jura.
We would like to thank Prof.\ Ben Zuckerman for useful discussion and comments on multiple
versions of the text presented here. We thank Prof.\ James Day for useful insight regarding
aluminum and Prof.\ Steve Desch for suggesting travertine and plate tectonics as a possibility. 
We thank the anonymous referee for suggestions that helped improve this paper.
C.M.\ was supported by grants from the National Science Foundation 
(AST-1313428) and NASA (NNX14AF76G and HST-GO-13700.001-A).
P.D.\ acknowledges support in part by the NSERC Canada and by the Fund FRQNT (Qu\'ebec).
This paper is based on observations made with the NASA/ESA $Hubble$ $Space$ $Telescope$, 
obtained at the Space Telescope Science Institute, which is operated by the
Association of Universities for Research in Astronomy, Inc., under NASA contract NAS 5-26555.
Some of the data presented herein were obtained at the W.M. Keck Observatory, 
which is operated as a scientific partnership among the California Institute of 
Technology, the University of California and the National Aeronautics and Space 
Administration. The Observatory was made possible by the generous financial 
support of the W.M. Keck Foundation.
The authors wish to recognize and acknowledge the very significant cultural role and 
reverence that the summit of Mauna Kea has always had within the indigenous Hawaiian 
community.  We are most fortunate to have the opportunity to conduct observations from 
this mountain.
This research has made use of NASA's Astrophysics Data System.



\facilities{HST (COS), Keck:I (HIRES)}

\software{MAKEE, IRAF, IDL}




}




\clearpage

\begin{figure}
 \hspace{-0.3in} 
 \includegraphics[width=200mm]{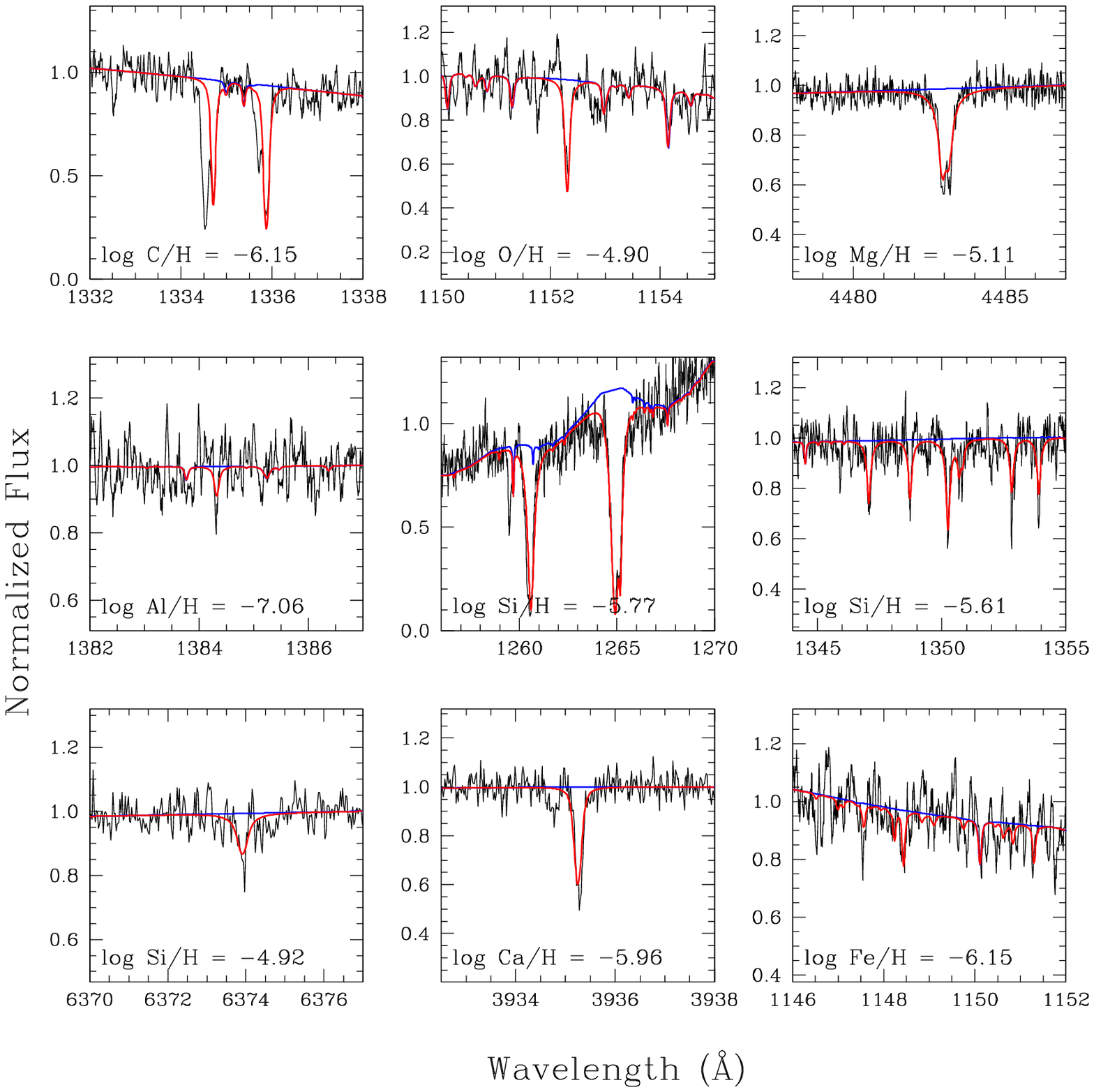}
 \vskip -0.1in
\caption{\label{figmetals} \large{Elements identified in the $HST$/COS and Keck/HIRES 
              spectra of SDSS\,J1043+0855.
              The black noisy curve is the data, smooth red curves are model fits, and 
              blue curves with the element
              of interest removed show which line or lines are being considered.
              Lines not included in the fit in the C panel are from interstellar carbon
              (see Section \ref{secspec}).
              The multiple panels for Si demonstrate the disagreement between
              optical and ultraviolet abundances; see discussion
              in Section \ref{secspec}. All wavelengths are in
              vacuum and presented in the heliocentric
              reference frame.  
              Measured abundances for the
              element of interest are indicated  
              relative to H by number in each panel.}  }
\end{figure}

\clearpage

\begin{figure}
 \centering
 \includegraphics[width=160mm]{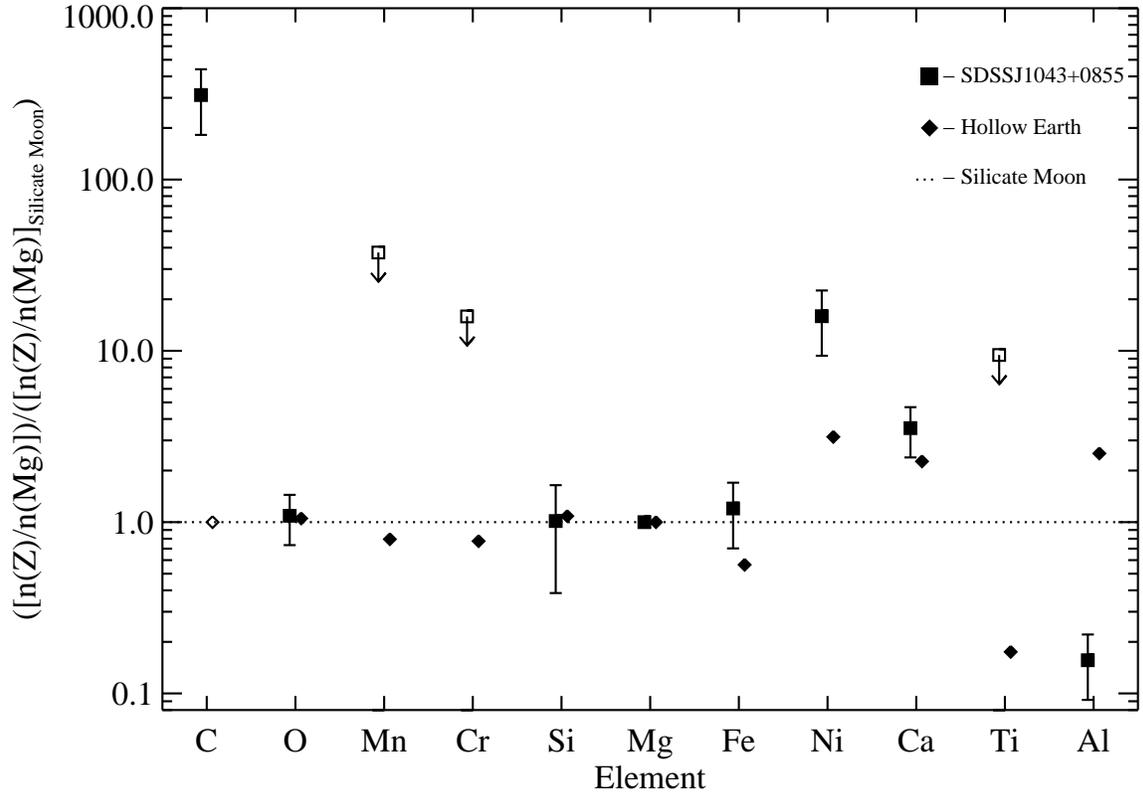}
 \caption{\label{fig1043z} \large{
               Abundances relative to Mg as compared to the silicate Moon (see Table \ref{tab1043});
               elements have increasing condensation temperature 
               (for physical and chemical conditions within the Solar nebula) to the right.
               The square data points are measurements and upper limits for SDSS\,J1043+0855
               from Table \ref{tab1043}, uncertainties are $\approx$50\% less than what would be expected
               by simple propagation of errors as discussed in \citet{klein10}. 
               The diamond data points are
               for a model where the core and lower mantle of the Earth are removed as described
               in Section \ref{secdisc}. 
               The dotted line is what would be expected if abundances relative to Mg exactly like
               the silicate Moon are realized. C measurements for the silicate Moon
               and the crust and mantle of the Earth are discussed in more detail in
               Section \ref{secdisc} and Table \ref{tab1043}.} }
\end{figure}

\clearpage

\begin{figure}
 \centering
 \begin{minipage}[t!]{80mm}
  \hspace{-0.5in}
  \includegraphics[width=105mm]{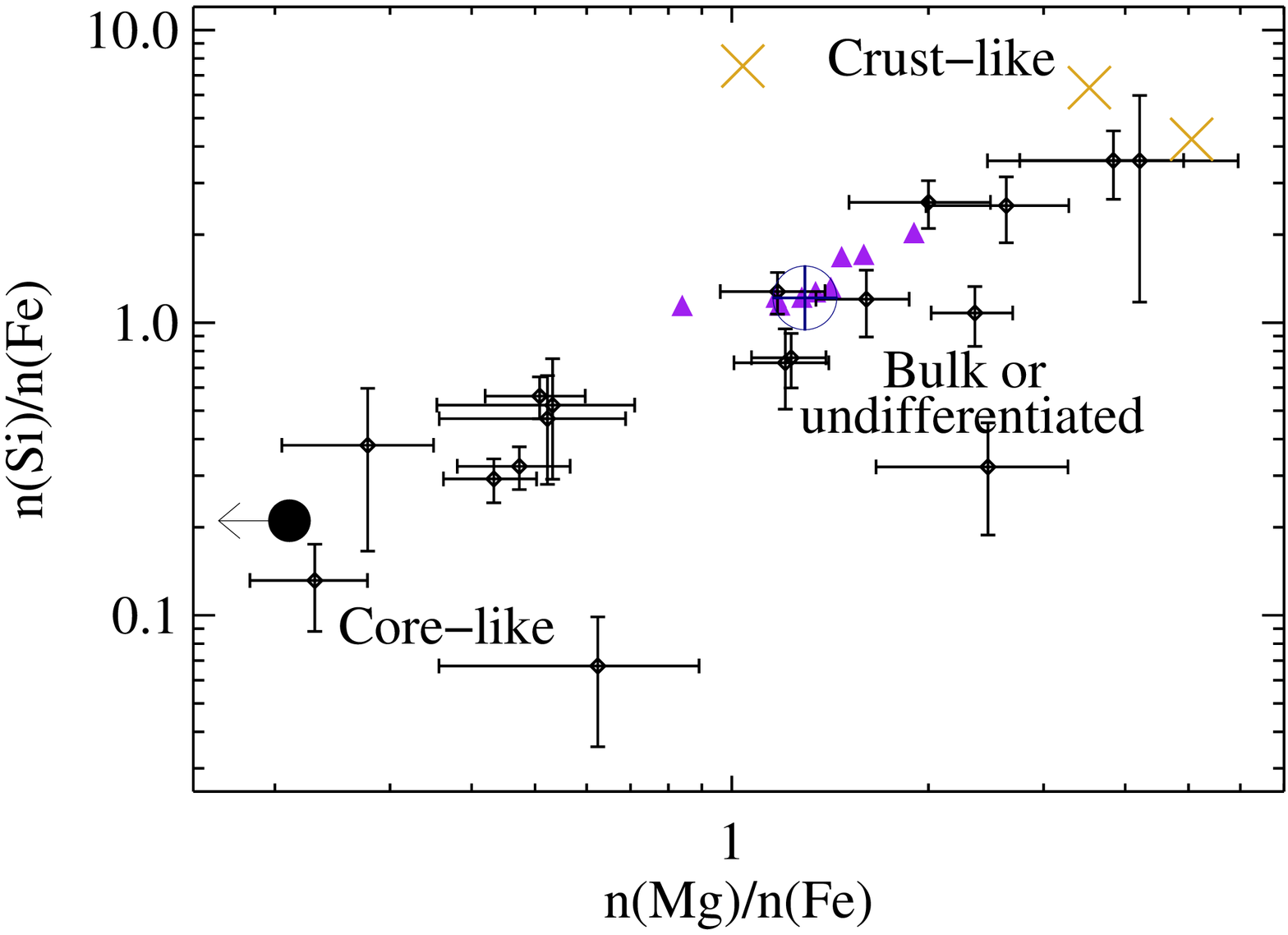}
 \end{minipage}
 \\*
 \begin{minipage}[b!]{80mm}
  \hspace{-0.5in}
  \includegraphics[width=105mm]{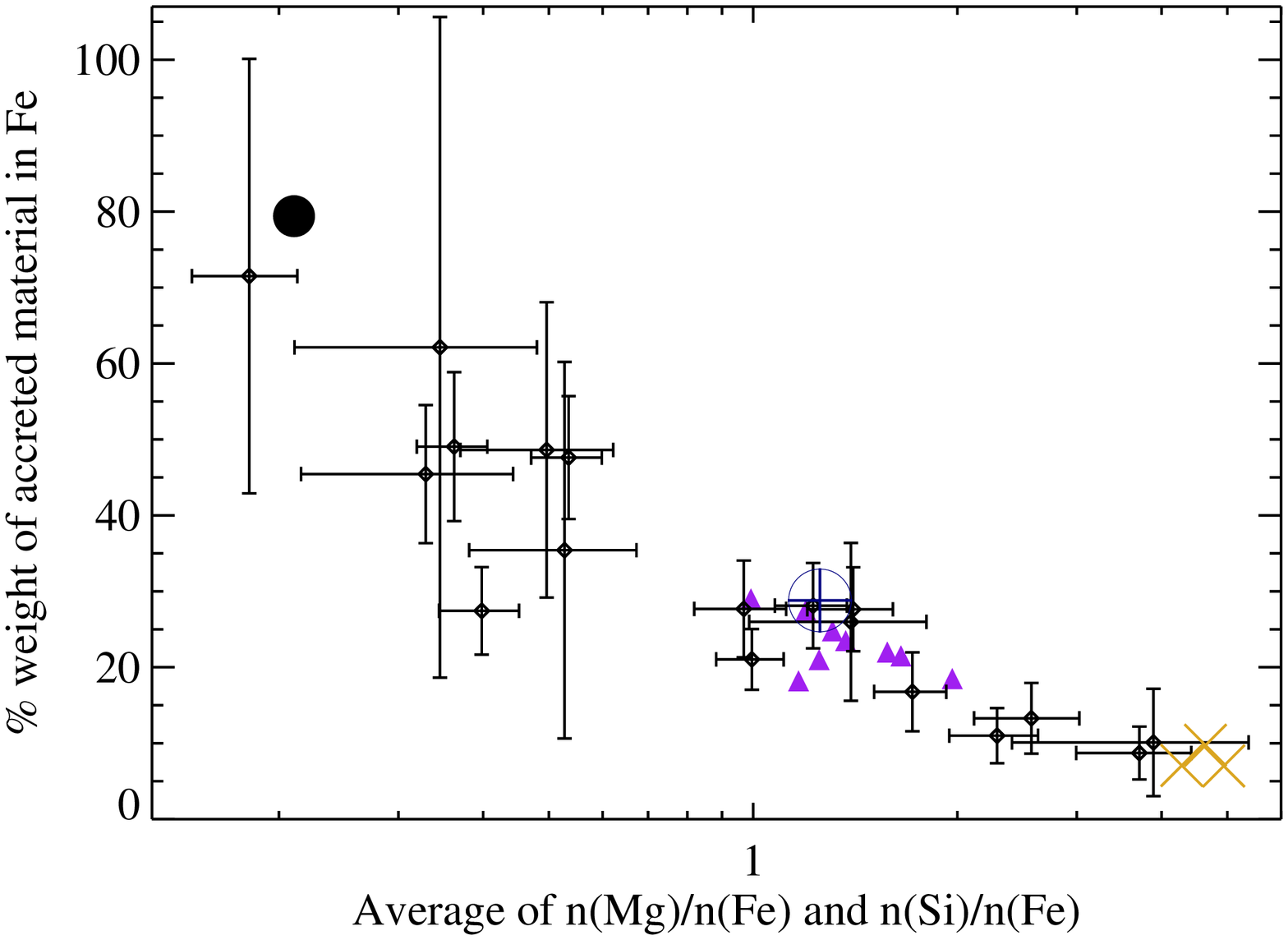}
 \end{minipage}
 \caption{\label{figtrend} \normalsize{
               A trend when comparing accreted Si/Fe vs Mg/Fe abundance ratios (top panel) is 
               suggestive of whether material accreted by white dwarfs originates from the inner 
               or outer regions of a rocky body. Those
               objects with high relative levels of Fe lie in the region of the plot 
               marked as ``Core-like'' as they are reminiscent of the Fe-dominated interior of 
               differentiated rocky bodies. Those objects with low relative levels of Fe lie in the region 
               marked as ``Crust-like'' as they appear significantly devoid
               of Fe as one might expect if the outer layers of a differentiated body were stripped and then
               accreted by the host white dwarf star. 
               The bottom panel confirms the Fe-rich or Fe-poor nature
               of objects in these two regions. The large, gold-colored ``X'' points 
               that lie in the ``Crust-like'' region mark the silicate Moon,
               bulk oceanic crust of the Earth, and bulk continental crust of the Earth.
               The silicate Moon has Mg/Fe$\approx$Si/Fe$\approx$5 and Fe mass
               fraction of $\approx$10\%
               while the bulk continental crust of Earth has 
               Mg/Fe$\approx$1, Si/Fe$\sim$8, and Fe mass fraction of $\approx$7\%.
               Small, filled purple triangle data points present chondrite data and are meant to represent
               extremely primitive rocky objects in the Solar system.
               The large, blue circle with a blue plus inside it near 
               Mg/Fe$\approx$Si/Fe$\approx$1.3 is the Bulk Earth.
               The large, black filled circle is for the core of Earth which is based on extrapolations
               from chondrite data and assumes no Mg is in the core \citep{allegre95}; 
               the Earth core value for Mg/Fe is set to Si/Fe for plotting purposes.
               White dwarf data come from 
                \citet{dufour10,dufour12}, 
               \citet{zuckerman11}, 
               \citet{gaensicke12}, 
               \citet{farihi13}, 
               \citet{xu13,xu14,xu16}, 
               \citet{wilson15},  
               \citet{raddi15}, and 
               \citet{farihi16}. 
               Metal abundance ratio errors for white dwarf measurements are adjusted similar to those
               for SDSS\,J1043+0855 like in Figure \ref{fig1043z}. Percent weight of Fe errors
               are as propagated directly from the Fe abundance measurement error in the atmosphere
               of each white dwarf star.
               Bulk Earth and Earth core data are from \citet{allegre95}, chondrite data from \citet{wasson88},
               silicate Moon data from \citet{lodders11}, and bulk oceanic and continental crust data from \citet{anderson89}.} }
\end{figure}


\clearpage

\floattable

\begin{deluxetable}{ccccccc}
\rotate
\tabletypesize{\large}
\tablecolumns{7}
\tablewidth{0pt}
\tablecaption{\large{SDSS\,J1043+0855 Metal Pollution} \label{tab1043}}
\tablehead{ 
  \colhead{$Z$} & 
  \colhead{log[$n$($Z$)/$n$(H)]$_{\rm measured}$} & 
  \colhead{log($\tau$$_{\rm diff}$/years)\tablenotemark{a}} &
  \colhead{[$n$($Z$)/$n$(Mg)]$_{\rm accreted}$\tablenotemark{b}} &
  \colhead{[$n$($Z$)/$n$(Mg)]$_{\rm Silicate\,Moon}$\tablenotemark{c}} &
  \colhead{$\dot{M}$$_{\rm acc}$/(10$^7$ g s$^{-1}$)\tablenotemark{d}} &
  \colhead{\% wt.\tablenotemark{e}}
}
\startdata
C    & $-$6.15$\pm$0.30  & $-$2.32 & 0.13       & 0.0004   & 0.15       & 1.2 \\ 
O    & $-$4.90$\pm$0.20  & $-$2.48 & 3.3       & 3.1   & 5.3       & 41 \\ 
Mg  & $-$5.11$\pm$0.20  & $-$2.16 & 1.0        & 1.0   & 2.4       & 18 \\ 
Al    & $-$7.06$\pm$0.30 & $-$2.23 & 0.013       & 0.084   & 0.035       & 0.27 \\ 
Si    & $-$5.33$\pm$0.50 & $-$2.31 & 0.84           & 0.84   & 2.3       & 18 \\  
P     & $-$7.40$\pm$0.30 & $-$2.41 & 0.0091       & 0.00006 & 0.028       & 0.22 \\
S     & $-$6.36$\pm$0.40\tablenotemark{f} & $-$2.54 & 0.13       & $-$         & 0.43       & 3.3 \\
Ca  & $-$5.96$\pm$0.20  & $-$2.36 & 0.22       & 0.064   & 0.90       & 6.9 \\ 
Sc   & $<$$-$7.5              & $-$2.41 & $<$0.0072 & 0.00003 & $<$0.032 & $<$0.25 \\
Ti     & $<$$-$7.0             & $-$2.44 & $<$0.024 & 0.0025  & $<$0.11 & $<$0.89 \\ 
V      & $<$$-$7.5             & $-$2.47 & $<$0.0082 & 0.0001  & $<$0.042 & $<$0.32 \\
Cr     & $<$$-$6.5             & $-$2.47 & $<$0.083 & 0.0052  & $<$0.43 & $<$3.3 \\ 
Mn    & $<$$-$6.5             & $-$2.52 & $<$0.094  & 0.0025  & $<$0.52 & $<$3.9 \\ 
Fe   & $-$6.15$\pm$0.30  & $-$2.58 & 0.23        & 0.20  & 1.3       & 10 \\ 
Ni    & $-$7.38$\pm$0.30 & $-$2.56 & 0.013        & 0.0008  & 0.078       & 0.60 \\ 
\enddata
\tablenotetext{a}{Diffusion constants, see \citet{koester09} for discussion of what these values mean and Section \ref{secdisc} for where these values come from for the analysis presented herein.}
\tablenotetext{b}{Parent body abundances relative to Mg; see Section \ref{secdisc} and Figure \ref{fig1043z}.}
\tablenotetext{c}{Silicate Moon data from Table 4.7 of \citet{lodders11} with the exception of data for C which comes from \citet{wetzel15}. See also Figure \ref{fig1043z} and discussion in Section \ref{secdisc} for C.}
\tablenotetext{d}{$\dot{M}$$_{\rm acc}$($Z$)=$M_{\rm env}$($Z$)/$\tau$$_{\rm diff}$($Z$) where $M_{\rm env}$($Z$) is the mass of element $Z$ in the hydrogen-dominated envelope of SDSS\,J1043+0855 assuming it has a mass of 2.82 $\times$ 10$^{16}$\,g (see Section \ref{secdisc}).}
\tablenotetext{e}{Mass fraction of each element in percent weight as deduced from the relative contribution to the total $\dot{M}$$_{\rm acc}$.}
\tablenotetext{f}{Sulfur is tentatively detected and the suggested abundance can be taken as a firm upper limit.}
\end{deluxetable}

\end{document}